\documentclass[10pt,twocolumn]{article}
\pdfoutput=1
\usepackage[latin2]{inputenc} 
\usepackage{a4wide} 
\usepackage{graphicx} 

\usepackage{tikz}
\usepackage[colorlinks=true, linkcolor=blue]{hyperref}
\usepackage{verbatim}

\makeatletter \newcommand \listoftodos{\subsection*{Todo list} \@starttoc{tdo}}
\newcommand\l@todo[2]
    {\par\noindent \textit{#2}, \parbox{10cm}{#1}\par} \makeatother

\definecolor{orange}{rgb}{1,0.5,0}
\tikzstyle{notestyle} = [draw=black, fill=orange, text width = 3cm]
\tikzstyle{notestyleleft} = [notestyle]
\tikzstyle{connectstyle} = [draw = orange, thick]
\tikzstyle{notestyle} += [text width=\marginparwidth]
\tikzstyle{notestyleleft} += [text width=0.9\marginparwidth]

\begin{document}



\hypersetup{pdfauthor={Petr Baudiš},pdftitle={Current Concepts in Version Control Systems}}

\title{Current Concepts in Version Control Systems}
\author{Petr Baudiš\footnote{The work on this paper was in part sponsored by Novell (SUSE Labs).}}
\date{2009-09-11}
\maketitle

\begin{abstract}
We give the reader a comprehensive overview of the state
of the Version Control software engineering field, describing
and analysing the concepts, architectural approaches and
methods researched and included in the currently widely used
version control systems and propose some possible future
research directions.
\end{abstract}




\section*{Preface}

This paper has been originally written in May 2008 as an adaptation
of my Bachelor
thesis and I~have been meaning to get back to it ever since, to add
in some missing information, especially on some of my research ideas,
describing the patch algebra more formally and detailing on the more
interesting chapters of historical development in the field.

Unfortunately, it is not likely that I will ever get the time to polish
the paper up to my full content; in the meantime, it is slowly getting
obsolete and I believe it can have a good value even as it stands,
especially as an introductory material for people getting into this
area. Maybe I~have missed some of the latest developments; I~ask
the reader for tolerance of the omissions, and my fellow
researches and authors for followups on this work that would fill
in the gaps and explain ongoing new ideas in the field.
\break
~\qquad---\quad Petr Baudiš

\section{Introduction}

Version control plays an important role in most of the creative processes
involving a computer. Ranging from simple undo/redo capabilities of most
office tools to complicated version control supporting branching and diverse
collaboration of thousands of developers on huge projects. This work will focus
on the latter end of the scale, describing the modern approaches for advanced
version control with emphasis on software development. We will compare various
systems and techniques, but not judge their general suitability as the set of
requirements for version control can vary widely. We will not focus on
integration of version control in the source configuration management
systems\footnote{Source Configuration Management covers areas like version
control, build management, bug tracking or development process management.}
since SCM is an extremely wide area and many other papers already cover SCM in
general. \cite{ossvcsstudy} \cite{scmconcepts} \cite{vmodels}

This work aims to fill the sore need for an
introduction-level but comprehensive summary of the current state of art in
version control theory and practice. We hope that it will help new
developers entering the field to quickly understand all the
concepts and ideas in use. Since the author is (was) a Git developer,
we focus somewhat on the Git version control system in our examples, but we
try to be comprehensive and cover other systems fairly as much as is within
our power and knowledge.

We also describe for the first time some techniques so far present only as
source code in various systems without any explicit documentation and present
some of our original work.
One problem in the field of version control is that since most innovation
and research happens not in the academia or even classical industry, but the
open community, very few of the results get published in the form of a
scientific paper or technical report and many of the concepts are known only
from mailing list posts, random writeups scattered around the web or merely
source code of the free software tools. Thus, another aim of this work is tie
up all these sources and provide an anchor point for future researchers.

\section{Basic Design}

The basic task of version control system is to preserve history of a collection
of files ({\it project})
so that the user can compare two historical versions
of the content easily, return to an
older version or review the list of changes between two versions.

\subsection{Development Process Model}

We will focus on version control systems that are based on the {\em distributed
development} paradigm. While the concept dates many years back, distributed
development has only recently seen major adoption and so far mostly only in the
area of open source development.

The classical approach to version control is to have a single central
server\footnote{Possibly with several read-only mirrors for replication, as in
the BSD ports setup.} storing the project history; the client can have
the current version cached locally, and possibly also (parts of) the history,
but any operation involving change in the history (e.g. committing new version)
requires cooperation with the central server.

This {\it centralized development} approach has two major shortcomings: first,
the developer has to have network access during development, which is problematic
in the era of mobile computing --- i.e., when developing during travel or fixing
problems at the customer site. Second, the central repository access implies
some inherent bureaucracy overhead --- the developer has to gain appropriate
commit permissions before being able to version-control any of his work,
development of third-party patches is difficult and since the repository is
shared, the developer is usually restricted in experimentation and creating
short-lived branches, often resorting to using some kind of manual version control
during regular workflow and using the VCS only on official occassions.

On the other hand, since all the development is tracked at a
central place, this model may pose advantages to corporate shops because work
and performance of all developers can be tracked easily.\footnote{In case of
in-company development using distributed version control, this can be
alleviated by setting appropriate policy for the developers to push their work
to designated central repositories in regular intervals.}

In the {\it distributed development} model, a copy of the project created by version
control has full version tracking capabilities and not only is all the history
usually copied over ({\it pulled}), but it is also possible to commit new
changes locally; later, they can be sent ({\it pushed}) to another repository in
order to be shared with other developers and users. This approach effectively
makes every copy a stand-alone repository with stand-alone branches; nevertheless,
most systems make it easy to emulate the centralized version control model as a
special-case at least for the read-only users.

\subsection{Snapshot-Oriented History}

There are two possible approaches to modeling the history. The first-class
object is usually the particular {\it version}\footnote{We use the terms
{\it version}, {\it revision} and {\it commit} interchangeably in this paper
based on the context, according to usual practice.} as a state ({\it snapshot}) of
the content at a given moment, while the changes are deduced only by
comparing two versions.  However there is also a competing approach which
focuses on the {\it changes} instead: in that case, the important thing when marking
new version is not the new state of the content, but the difference between the
old and new state; a~particular version is then regarded as a certain
combination of changes.\footnote{Note that the conceptual model may have no
relation to actual storage, which can well use deltas even within snapshot-oriented
systems.}

\subsubsection{Branches}

With regard to the snapshot-oriented history model, it may suffice in the
trivial case to merely represent the succeeding versions as a list with the
versions ordered chronologically.  But over time it may become feasible to work
on several versions of the content in parallel, while also recording further
modifications using the version control system; the development has effectively
been split to several variants ({\it branches}).

Handling branches is an obvious feature of a version control system as the
necessity to branch the development is frequent for many projects targetted at
wider audience. Commonly, aside of the main development branch another branch
is kept based on the latest public release of the project and it is dedicated to
bugfixes for this release.

Moreover, branches have many other uses as well --- it may be feasible for
individual developers or teams to store work in progress yet unsuitable for the
main development branch in separate branches, or even for a developer to save
incomplete modifications during short-time reassignment to a different task or
when moving to another machine. The prerequisite for this is that creating a
separate branch must have only minimal overhead, both technical (time and space
required to create a new branch) and social (it must be possible to easily
remove the branch too; also, the unimportant branches should not cloud the view
of the important major ones).

The ability to properly record and manipulate branches nevertheless becomes
truly vital as the development gets distributed, since in effect every clone
of the project becomes a separate branch. If the system aims to fully and
comfortably support the distributed development model, it must make sure that
there are no hurdles for creating many new branches; furthermore, there can be
no central registry of branches, thus the branch-related data structures must
allow for branches to grow from different points in history with full independence
and dealing with naming conflicts,\footnote{In independent repositories, branches
and revisions can be given conflicting names, as will be covered in detail later.}
shall they ever meet again in some repository.

Aside of the obvious operations with branches (referring to them, comparing
them, storing them in a reasonable manner, switching between them), perhaps the
technically most challenging one is merging --- combining changes made
separately in two branches in a single revision. In the past, only the most crude
methods for merging two branches were available, but again with the onset of
distributed version control this aspect has grown in importance (a merge
essentially happens every time a developer exchanges modifications with
the outside world). We will explain the challenges associated with
merging in section \ref{merging}; one important consideration we
shall mention now is that many of the merging techniques require to find the
latest point in history common to both of the to-be-merged branches (or, more
generally, which changes are not common to the history of both branches).

Aside of merging, another desirable operation is so-called {\it cherry-picking}:
taking individual changes from one branch to another without merging them
completely. While most modern systems support this in a rudimentary way, the
operation is not natural in snapshot-oriented systems and can result in trouble
when merging the branches later, both in duplicated changes stored in the history
and increased merge conflicts rate.\footnote{Git provides a {\bf git-rerere} tool for
automatically resolving merge conflicts that have been resolved in a certain
way before, thus somewhat reducing this problem.}

\subsubsection{Data Structures}

Given that the development may fork at some point in history, we cannot use
simple lists anymore to adequately represent this event but we have to model
the history as a directed tree instead, with succeeding versions connected by
edges; the node representing a {\it fork point} (not only one, but several
branches grow from the node) has multiple children.

However, when we take merging into consideration, it becomes necessary to also
represent the past merges between branches if we are to properly support
repeated merges and preserve detailed history of both branches across the
merge. The commonly used method is to use a more general {\it directed acyclic
graph} (DAG; figure \ref{fig:dag} several pages later) to represent the history;
compared to a directed tree, DAG lifts the restriction for a node to have only
a single parent. Thus, versions representing merges of several branches have the
last (``head'') versions of all the merged branches as their parents.  The
acyclic property makes sure that a commit still cannot be its own ancestor,
thus keeping the history sane (time travel and quantum effects
notwithstanding).

An interesting problem to note is how to actually define a branch.
In case of trees, the branch is defined naturally by the tree structure,
however in a DAG this becomes less obvious. One approach is to record branch of
each revision explicitly (e.g. Mercurial), another is to give up on sorting out
revisions to explicit branches altogether (e.g. Git) --- in that case, a
``branch'' is merely a named pointer to the commit at the head of the branch,
with no clear distinction of which ancestors belong to the same branch; it
might be tempting to keep branch correspondence through fixed parent nodes
ordering in merge nodes (``first parent comes from the same branch''), however
this approach has caveats as described in \ref{fastforward}.

\subsection{Changes-Oriented History}
\label{cset}

Alternatively to snapshot-oriented history, a~ra\-di\-cal\-ly different approach is
to focus on the changes as the first-class object --- the user creates and
manipulates not a~particular version of the project, but instead a~particular
{\it change}; a~``version'' is then described simply by given combination of
changes recorded.  The major advantage is that merging operations may be much
more flexible, and inherently more data is preserved as the changes themselves
are recorded and do not have to be deduced a posteriori from checking the
difference between two snapshots.

The main representative of this class is {\bf Darcs}
(see \ref{darcs}) with its main trait of the powerful formalisation {\it patch
algebra} (see \ref{palg}) which makes it possible to easily perform merging of
two versions by adding the changes missing in the other version and reordering
the changes in a well defined way.

In order to make full use of the smart merging techniques like patch algebra,
in some cases the operation time can degrade exponentially to the number of
revisions in the current implementations \cite{darcsexp};
making the algorithms more efficient is
subject of active research, however at least until very recently,
practicality of changes-oriented history model was severely limited and this
approach was not suitable for large histories; we are not aware of performance
improvement studies for the recent Darcs improvements and it is uncertain
if these complexity issues are inherent to the whole changes-oriented model.

\subsection{Objects Identification}

The most visible (and diverse) aspect of the history models is the problem of
identifying the history objects (snapshots or changes, depending on the model).

The old {\bf RCS} and {\bf CVS} tools use a simple scheme where every
version is identified by a string of dot-separated numbers; the first number is
usually 1, the second number is the version number on the main branch. If it is
required to identify a version on a different branch, {\tt
BASE\_VERSION.BRANCH.VERSION} scheme is used --- {\tt BRANCH} is the number of
the branch forked from the base version and {\tt VERSION} is the sequence
number of the given version on the branch; this scheme can be used recursively
to identify versions on branches of branches etc.

However, we shall remember that in a distributed system, we cannot see all the
used identifiers in all repositories (since they can be on
disconnected systems), thus we meet with special
problems when having to simultanously assign new identifiers in different
disconnected repositories.  Eventually, we are met with three conflicting
requirements\footnote{For clarity, in this
context we use ``simple'' instead of ``memorable'' and ``stable'' instead of
``global'' presented in the literature.
The latter correspondence is somewhat loose and given under the
assumption that it does not make sense to use central authority for assigning
identifiers in case of distributed version control --- thus, either the
identifiers are globally-meaningful and stable, or they have only local meaning
and thus inter-repository collisions are inevitable and relabeling in that case
necessary.} \cite{names}\cite{petname}:

\begin{itemize}
\item {\it Uniqueness:} The identifier is guaranteed to be unique and always
identifies only a single history object in a given repository.
(Note that this requirement is frequently slightly compromised by only
hoping that the identifier is unique with a very high probability.)
\footnote{Here we mean uniqueness guaranteed by technical means,
preferrably provably secure.
E.g., GNU Arch does use identifiers that are ``unique'', but only by
policy as one part of the id is user's e-mail address --- these are unique,
but there is no technical safeguard that different users will not input
the same e-mail address; this approach requires trusting the other
repository database to contain correct content.}

\item {\it Simplicity:} The identifier is easy to use --- reasonably short
and preferrably informative to the user (e.g. giving some information
on ordering of the revisions).

\item {\it Stability:} The identifier does not change over time and
between different repositories.
\end{itemize}

The requirements are of conflicting nature --- we can generally always pick only two and
sacrifice the third. Different version control systems make different choices
here: for example, Git, Mercurial and Monotone sacrifice some
usability and assigns every commit a 40-digit SHA1 hash \cite{SHA1} depending on the
current content and history of the version.\footnote{This also technically
sacrifies some uniqueness, but the probability of collision of two hashes is so
small that it is deemed to be unimportant for practical purposes. This has been
disputed by \cite{Henson} \cite{Henson2}, however that paper has been criticized by
the version control \cite{mthash} and cryptographic \cite{Black} community.}
To alleviate the UI problem somewhat, the systems allow shortcutting
the long hashes to only few leading digits, as long as they match uniquely.

BitKeeper and Bazaar\footnote{Technically, Bazaar internally uses
globally unique identifiers, but RCS-style identifiers are the primary way
of user interaction.} sacrifice the stability and use RCS-style
identifiers that are unique only within a single repository and can shift
during merges.
Darcs sacrifies the uniqueness and uses user-assigned
identifiers for individual changes (however, the identifiers are encouraged
to be rather elaborate, so collision probability should not be very high).

\subsection{Content Movement Tracking}
\label{moves}

One interesting problem in version control is following particular content
across file renames, copies, etc. Practically all systems offer a way to list
all revisions changing given file or set of files, however things become more
troublesome when a file is renamed or copied or when we want to check history
of finer-grained objects than whole files. Most systems support so-called
``annotated view'' of a file, for each line showing the commit that introduced
it in its current version. However, this operation is frequently very expensive
and on many occassions the line-level granularity is not suitable either.

While old systems like CVS do not support tracking file renames at all and the
new files start with fresh history unless the repository is manually
modified,\footnote{The modification usually involves manually copying the file
under new name within the repository database. However, this presents problems when the
working tree is seeked to an older point in the history or in case of
branches.} most newer systems support file renames, usually by providing a
special command that records the movement --- less common approach used e.g. by
GNU Arch is to autodetect movement by tracking unique tags within the objects
(strings within files or hidden files in directories).

In that case, showing history of the new file will usually automatically switch
to the older file name at the point of the rename, and there is also a
possibility of cross-rename merges when a file has been renamed between the two
revisions.  However, this carries e.g. the danger of making the merging
operation complexity proportional to project history, something many systems
seek to avoid. Also, as the support gets naturally extended to recording not
only renames but also file copies, it becomes much less obvious which file
instances to merge; another natural extension to record file combinations
makes the situation even more challenging.\footnote{The underlying
design of recording renames varies --- for example Subversion records
renames as copy+delete operations, which can bring some inherent problems e.g.
for using the rename information during merges. \cite{hudsonc}}

A somewhat controversial alternative approach has been taken by Git,
where file renames and copying is not explicitly recorded with the rationale
that Git aims to track content on a more fine-grained level and hard-coding
renames information would pollute the history with potentially bogus
information. \cite{gitsct} Instead, Git can heuristically detect the events by computing
similarity indexes between files in the compared revisions,\footnote{The
similarity index is computed by hashing line-based file chunks of the files and
comparing how many of the hashes exist in both
files. \cite{gitsrcdiffcoredelta}} and it provides a so-called ``pick-axe''
mechanism to show changes adding or removing a particular string. Thus, when
feeding pick-axe with some chunk of code, it will show the commits that
introduced it to the files, as well as commits that removed it from files,
effectively giving the user a rudimental ability to trace the code history.\footnote{An
obvious idea for an extension is to have the GUI frontends integrate mouse
text selection with pickaxe, further simplifying the process.}

The problem of showing history of particular parts of files and following
semantic units across files is still open and will in our opinion benefit from
further research (as elaborated in \ref{futuremove}).

\section{Current Systems}

We have already mentioned several version control systems when describing basic
design aspects; now, we shall briefly describe the most widespread and
interesting ones in more detail. We will focus on systems that are freely
available and in use in the open-source community, and even so not go through
all of them but instead highlight the most influential ones.
We will not cover most proprietary systems like ClearCase or SourceSafe in detail
since we feel they generally do not bring much innovation into the version control
field itself (focusing more on process management, integration with other
software etc.) and detailed information is hard to gather from open sources.

Up to CVS, the evolution was rather slow, but then with growing need for
distributed version control, many short-lived projects sprang up and explored
various approaches; now the field has rather stabilized again, with three major
systems currently used: Bazaar, Git and Mercurial.

\subsection{The Two Grandfathers}

The first true version control ever in use was {\bf SCCS} (Source Code Control
System) \cite{SCCS}, introduced in 1972 at Bell Labs and distributed commercially. It uses
the weave file format (\ref{weave}) to track individual files seprately and
provides only a rather arcane user interface.  Although its use in the industry is very
rare nowadays, the weave storage format has resurfaced recently.

A competing system {\bf RCS} (Revision Control System) \cite{RCS} published in
1985 has seen much more success, even though it also operates only on isolated files.
It uses a delta sequence file format (\ref{rcsdelta}) and the rapid adoption rate
compared to SCCS is likely to be accounted mainly to much more user-friendly interface
(though it is still rather rudimentary by current standards). RCS is still
sometimes used in present for version-controlling individual files, e.g. when
maintaining system configuration. However, its concepts and file format
has seen even much more widespread adoption thanks to CVS.

\subsection{Concurrent Version System}

{\bf CVS} \cite{CVS} is essentially an RCS spinoff, bringing two major extra capabilities:
ability to version-control multiple files at once and support for parallel
work, including network support.
It has seen extremely widespread adoption and has long remained the de-facto
standard in version control, at least among the free systems.

However, CVS is directly based on RCS, and it is ridden with inherited artifacts of
individual file revision control --- the need for elaborate
locking, difficult handling of branching and merging and lack of commits
atomicity: when committed change spans multiple files, the commit will
be stored for each file separately and there is no reliable way to reconstruct
the whole change.\footnote{However, relatively reliable heuristics are implemented, e.g. by
the {\bf cvsps} project. \cite{CVSps}}

\subsection{Modern Centralized Systems}

{\bf Perforce} \cite{perforce} is a widely commercially used proprietary
version control system that provides rather advanced version control
features, though it keeps within the centralized model. Compared to CVS,
it supports atomic commits (tying changes in multiple files together),
tracking file renames and more comfortable branching and merging support
(branches are presented as separate paths in the repository and repeated
merges are supported).

In the open-source version control realm, {\bf SVN} \cite{SVN} has been
developed as a direct CVS replacement and currently appears to remain
the final step in the evolution of centralized version control. It shares many
of the Perforce features, though many workflows are different. It is seeing
rapid adoption \cite{svnsurvey} and basically dethroned CVS as the default choice of centralized
version control for open-source projects.

\subsection{Older Distributed Systems}

{\bf BitKeeper} \cite{bk} is another proprietary version control system worth
mentioning, since it pioneered the distributed version control
field\footnote{The concept of distributed version control has been explored
before by e.g. Aegis, but BitKeeper was the first system to bring it to
widespread use.} and also has relatively well-understood capabilities and
internal working as it was for long time available for free usage by open
source projects and some prominent projects like the Linux Kernel used it.

BitKeeper supports atomic commits (though files have revisions tracked
individually as well), generic DAG history and the basic distributed version
workflows;\footnote{That is, the pull/push operations and good merging support.}
however, each repository contains only one branch.  RCS-style
revision numbers are used, same revisions can have different numbers in
different repositories and revision numbers on branches can shift during
merges.  Internally, BitKeeper uses the weave file format and makes full
use of its annotation capabilities in the user interface and during merges
(see \ref{pcdv}).
It is still actively developed and reportedly commercially popular.

With regard to the history, we shall also mention {\bf GNU Arch} \cite{arch} as
the first open-source distributed version control system that has seen wider
usage, however it suffered by bad performance and difficult user interface; the
project has been basically discontinued (though still formally maintained), but
Bazaar (see below) can trace back its ancestry to GNU Arch to a certain degree;
Darcs has been also somewhat inspired by its changeset-oriented design.

\subsection{Monotone}

Another important milestone is {\bf Monotone} \cite{mtn} with its design built
around a graph of objects identified by hash numbers computed from their
contents\footnote{Technically, the first system using hashes as identifiers is
{\bf OpenCM} \cite{opencm}, but this system has never really taken off as far
as we can tell.} serving as direct inspiration for Git and Mercurial. Monotone
itself puts strong emphasis on security and authentication, using RSA
certificates during communication, etc.  It is actively developed but does
not see as wide adoption as the three systems below, mainly due to
performance issues.

Git and Mercurial build upon the Monotone object model, which is generally
built as follows (Monotone implementation details notwithstanding):
There are three main object classes --- files (or blobs), manifests (or
directory trees) and revisions (or commits/changesets). All objects are
referenced by large-size hash of their content that should have guarantees akin
to cryptographical hashes (currently, SHA1 is used in all popular systems).
File objects contain raw contents of particular file versions.
Manifest objects contain listing of all files and directories\footnote{Git
uses one object per directory instead; there is a trade-off between object
reuse and dereferencing overhead in this choice.}
at a particular point of history, with references to objects with
appropriate content. Revision objects form the history DAG as a subgraph of the
general object graph by referencing the parent revision (or revisions, in case
of merge) and the corresponding manifest object. Usually, log
message and authorship information is also part of the revision object.

The reader should notice that this scheme is {\it cascading} --- by changing a
single file, new manifest object with unique id will be generated (the file
hash inside the manifest object changed, thus manifest object hash will change
too), and this will in turn result in a revision object with unique id as well.
Even committing the same content with the same log message in different
branches will not create an id clash since the parent revision hash will be
different within both objects. Tagging can be realized by simply creating named
reference to either the revision object or another object containing the
reference and also some tag description, possibly digitally signed by the
project maintainer to confirm that the tagged revision is an official release;
thanks to the usage of cryptographically-strong hashes for references, the
cryptographic trust extends to the complete revision content.

\subsection{The Three Rulers}


In spring of 2005 a rapid series of events resulted in termination of the
permission to use BitKeeper for free software projects, providing an immediate
impulse for improvement in the area of open-source version control systems, especially to
the Linux Kernel community --- almost in parallel, the Git and Mercurial
projects were started. Independently, Canonical Ltd. (the vendor of popular
Linux distribution Ubuntu) started the Bazaar\footnote{At that time, it has
been called BazaarNG --- a successor to an original Baz(aar) project which was
itself a fork of GNU Arch.} project around the same time.

All three systems provide atomic commits, store history in generic DAG
and support basic distributed version control workflows.

{\bf Git} \cite{git} \cite{gitsct} re-uses the generalized Monotone object model; by default
it stores each object in a separate file in the database, thus having a very
simple and robust but inefficient data model;\footnote{This scheme precludes
need for any locking whatsoever, but there is no compression in use and
using many small files means inherent overhead on the filesystem level.}
however, extremely efficient
storage using Git Packs (\ref{gitpacks}) is also provided.  In its user interface, Git
emphasizes (but does not make mandatory) the rather unique concept of
{\it index},\label{gitindex} providing a staging area above the working tree
for composing the next commit.\footnote{Originally a low-level mechanism, the
popularity of index usage among developers elected it to a first-class user
interface concept. If users choose to use it, they manually add files (or even
just chunks of changes) from the
working tree to the index at various points and then the index is committed,
regardless of working tree state; thus, the user may modify a file, add it to
index and then before committing modify it again for example with the changes required to
compile it locally --- the commit operation will use the older version of the file. When
explicitly recording all files to commit in the index, greater discipline of
making appropriately fine-grained commits can be encouraged. Also, non-trivial
conflicts are recorded in the index, making for a natural way of resolving them
by simply adding the desired resolution to the index.} Git is written in C
(and a mix of shell and Perl scripts) and is tightly tailored to POSIX systems,
which historically meant portability problems especially with regard to
Windows.  It is designed as a UNIX tool, providing large number of high- and
low-level commands, and it is traditionally extended by tools calling the
commands externally.

{\bf Mercurial} \cite{hg} \cite{revlog} has very similar object model to Git and Monotone, but uses the
revlog format (\ref{revlogs}) for efficient storage. It is written mainly in Python and thus does not
suffer from the portability problems of Git; since it uses an interpreted
language, the main extension mechanism is writing tightly integrated plugins.
Performance-sensitive portions are implemented in C.

While both of the systems above always emphasized their robust data models and
performance, {\bf Bazaar}'s \cite{bzr} main focus is simple and easy user interface;
though internally using elaborate unique revision identifiers, it presents
primarily the RCS-like revision numbers to the user, and its user interface has
direct support for some common workflows like emulation of the centralized
version control setup {\it (checkouts)} or {\bf Patch Queue Manager}, an
advanced alternative to all developers pushing to a central repository.\footnote{Developers
send their pull requests or actual patches to a public mailing list, where
they can be acknowledged or vetoed by others and then a robot automatically
picks them up, checks if they pass the required testsuites and eventually
merges them to the main branch. \cite{bzr-pqm}}
Bazaar is also written in Python, making the same tradeoffs as Mercurial.

\subsection{Darcs}
\label{darcs}

{\bf Darcs} \cite{darcs} is a somewhat unusual specimen in the arena of version control
as it focuses on changes as the first-class objects instead of trees; any
particular version of the repository is merely a combination of patches on
an empty tree, and manipulating the patches is the primary focus.

Darcs supports single branch per repository and names the patches based on user
input; particular combinations of patches can be tagged to mark specific
versions of the tree.\footnote{In fact, a Darcs tag is itself a patch that is
however empty and depends on all patches currently applied.} Darcs does not
organize the patches in a particular graph, they are structured more like a
stack, though a graph could be imagined based on interdependencies between the
patches; merges of several patches create auxiliary patches representing the
merges.

A particularly interesting feature of Darcs is precisely how it handles patch
merges --- it uses a formal system called {\it patch algebra} (\ref{palg})
to describe dependencies between patches and basic
operations on the patch stack required to properly merge two sets of patches
(branches).

Darcs is written in Haskell, which is somewhat exotic choice, but is well
portable. Unfortunately, as far as we are aware it suffers from severe
performance problems (as discussed in \ref{cset}) when used on large
repositories with long histories; more efficient merging algorithms as well as
proving formal correctness of the underlying patch theory is subject of ongoing
research.

\section{Storage Models}

The storage model of a system can be quite different from the logical design in
order to accomodate for practical considerations of space and time effectivity.
Here we shall explore possible approaches to permanent storage of the data and
metadata tracked by version control. All version control systems differ in
implementation details and various individual tweaks; we will not try to be
exhaustive here and will focus only on the most widespread and most interesting
ideas.

First, we will dwell on the currently popular algorithms to compare two objects
and create a {\it delta}: complete technical description of their
differences.\footnote{E.g.  a series of insert--copy commands or a ``diff''
tool output text.}
Second, we shall elaborate on possible ways of representing an individual delta
as well as whole single-file history, then we will look at the approaches on
organization of the whole repository.

\subsection{Delta Algorithms}
\label{delta}

There is a certain variety of delta algorithms in use, and there can even be
multiple algorithms used within a single system --- some algorithms provide
suitable and sensible output for human review while others are focused on
finding the smallest set of differences and providing minimal required output
for storage.

\subsubsection{Common Techniques}

First, we shall describe several common techniques used when dealing with
deltas. The first technique worth mentioning is used to our knowledge by all
commonly used systems newer than CVS: {\it delta combination} \cite{deltanotes}
\cite{revlog}. When applying a chain of deltas, they are first merged to a
single large delta, and only then applied.  This reduces the number of data
copies from $O(deltas)$ to $O(1)$ and takes advantage of cache locality,
dramatically increasing performance of applying delta chains.

Another interesting technique used most notably in Subversion is using {\it
skip-deltas} when storing delta chains. \cite{deltanotes} In order to limit
chain lengths, the deltas aren't always created against neighboring revision
but instead construct a binary tree alike structure --- thus, reconstructing
particular revision does not require $O(revs)$ deltas but only $O(\log revs)$.
However, when adding new revisions, parts of the tree or the whole tree needs
to be reconstructed, and the deltas are usually larger. Newer systems usually
solve the problem of too long delta chains by simply inserting the whole
revision text when the chain becomes too long or large in size.

\subsubsection{Myers' Longest Common Subsequence}

Perhaps the most widely used algorithm was proposed by Eugene Myers in \cite{Myers}
and is based on recursively finding the longest subsequence of common lines
in the list of lines of the compared objects; it is used most notably by the
{\bf GNU diff} tool~\cite{diffutils}, but also as the user-visible diff algorithm of
most systems.  The main idea of the algorithm is to perform two breadth-first searches
through the space of possible editations (line adds/removals/keeps), starting
from the beginning and end of the object in parallel; when the two searches
meet, we have found the shortest editation sequence.
The algorithm has quadratic computational complexity, so it is
unsuitable for very large inputs.

As noted in~\cite{revlog}, while this algorithm is quite suitable for human
consumption and does produce optimal longest common subsequences, it does not
produce optimal binary deltas since it weights additions, changes and removals
all the same, while binary deltas need to actually record only added data and
removed data does not need to be spelled out.\footnote{However, this should be
easy to work around by adjusting the breadth-first search to add 0-length
edges at the beginning of the queue and 1-length edges at the end.}

\subsubsection{Patience Diff}

An interesting variation of the above algorithm was recently introduced as the
default diff method to the Bazaar system, devised by Bram Cohen.
\cite{bzrpdiff} A common problem with using the Myers' algorithm is that the
common subsections search misaligns additions of parentheses on separate lines
or other simple syntactic constructs, making it harder to view by humans.

The main idea in this algorithm is to ``take the longest common
subsequence on lines which occur exactly once on both sides, then recurse
between lines which got matched on that pass.'' \cite{cdiff} Also, it achieves
better computational complexity by using Patience Sorting \cite{psort} to find
the longest common subsequence.\footnote{The referenced paper describes finding
the longest increasing subsequence. To find longest common subsequence, a
sequence of lines corresponding to one object is created, with the elements
being line numbers of the same lines in the other object; lines that are
present more than once or never are ignored in the search.}

\subsubsection{BDiff}
\label{bdiff}

The bdiff algorithm \cite{revlog} improving the Ratcliff--Obershelp pattern
recognition \cite{gestalt} comes from the Python {\bf difflib} library
\cite{pydifflib} and is used by Mercurial, both for delta storage and providing
difference listing to the user. In contrast to looking for the longest common
subsequence, it searches for the longest common continuous substring within the
objects and recursively in the parts preceding/succeeding it. It is quadratic
like the Myers algorithm, but has been measured \cite{revlog} to have better
average performance and output more human-friendly deltas.

\subsubsection{XDelta}
\label{xdelta}

The tool {\bf xdelta} \cite{xdelta} \cite{xdeltap} popularized use of the Rabin
Fingerprinting technique \cite{Rabin} for delta generation: it is effectively a
simple modification of the Rabin--Karp substring matching algorithm, inspired
by the work on {\bf rsync} \cite{rsync}.
Subversion \cite{svndiff}, Monotone \cite{mtnfaq} and Git
\cite{gitsrcdiffdelta} use this variation for internal storage ---
generating one-way binary diffs between
arbitrary non-textual blobs.

The algorithm produces {\sc Copy} and {\sc Insert} instructions on the output
and depends on fast hashing. The first object is divided into small
blocks\footnote{The window size is 16 bytes in Git; small power of 2 is
recommended in general.} and each block is hashed and added to a hash table
(proportional to input size).  Then, a running hash of the same-sized window is
being computed on the second file; the moment a match is hit, all found blocks
of the first object are stretched to largest possible matches to the current
position in the second object and the {\sc Copy} instruction is generated for
the largest match. {\sc Insert} instruction is generated for unmatched data in
second object.

This algorithm is very fast (having much better computational complexity
compared to Myers' --- $O(n)$ vs $O(n^2)$) and produces efficient diffs for
arbitrary data, however it is entirely unsuitable for human consumption.

A similar scheme is also used in zdelta \cite{zdelta} with reportedly better
efficiency and with code released under a more permissive licence.  ZDelta
reuses the LZ77 algorithm \cite{LZ77} for finding the matching parts and Huffman codes for
encoding the result.

\subsection{Delta Formats}

\subsubsection{Trivial Approach}

Not really a {\it delta} format by itself, the most naive approach to store
history is to simply store every version of every object separately, possibly
using some compression method (not taking advantage of historical context of
the object). This is the obvious format used for ``poor man's version control''
methods like archiving compressed snapshots of the whole project at regular
intervals and backing them up.  However, perhaps surprisingly this method does
see usage in some modern systems as well --- namely, it is used as the basic
in-flight storage format in Git.

\begin{figure*}
\begin{verbatim}
diff --cc git-am.sh
index 75886a8,4dce87b..b48096e
--- a/git-am.sh
+++ b/git-am.sh
@@@ -9,9 -8,9 +9,9 @@@ git-am [options] <mbox>|<Maildir>..
  git-am [options] --skip
 -d,dotest=       use <dir> and not .dotest
 +d,dotest=       (removed -- do not use)
  i,interactive   run interactively
- b,binary        pass --allo-binary-replacement to git-apply
+ b,binary        pass --allow-binary-replacement to git-apply
  3,3way          allow fall back on 3way merging if needed
\end{verbatim}
\caption{Combined diff example. \cite{gitmanual} The {\tt diff} and {\tt index}
lines are Git-inserted metadata, and the trailing text at the
{\tt @@@} line is only to give the reader a semantic context of the hunk.}
\label{v:combdiff}
\end{figure*}

\subsubsection{Unified Diff}
\label{unidiff}

The unified diff format \cite{unidiff} --- albeit not used in the changes
storage itself --- is the de-facto standard for interchange of plaintext changes, being
the usual format used in ``patch'' files.\footnote{Sometimes, a very similar
format called ``context diff'' is also used, however unified diffs are much
more common.} For each file, the diff contains all the {\it change chunks} ---
the areas of the file where changes occured. Each chunk consists of the chunk
header localizing the chunk in both versions of the file (starting line number
and line length of the chunk), {\it context} lines immediately surrounding the
changed lines (preceded by a space) and the change itself: added lines preceded
by a {\tt +} sign and removed lines by a {\tt -} sign; modified lines are written as a
{\tt +}/{\tt -} line pair. Unidentified lines in the diff are ignored, which allows
version control systems to include custom metadata in the diff.\footnote{Commonly,
informational-only notes about the compared revisions are inserted. However, e.g.
Git uses this to extend the diffs with file renames/copies information and can
make use of the metadata when applying the patches.}

This format has number of advantages: it is easy to generate, it is easy to
inspect by a human for the purpose of change review (including easily tweaking
details of the change in the unified diff itself), it is easy to apply
automatically\footnote{UNIX provides a standard {\tt patch(1)}
tool. \cite{POSIX}} and very importantly, it is possible to apply the
differences to another version than the one the diff was created from.  This is
allowed by the presence of the context lines: even if the starting line numbers
in the chunk header do not fit the context, the program can automatically
search the vincinity of the area for the context lines and apply the patch
fuzzily. At the same time, the context lines provide a way to verify that the
change is still applicable as-is to the target version.

We should also mention a Git extension to the unified diff for showing diff
representation of merges --- {\it combined diff} \cite{gitmanual} (see figure
\ref{v:combdiff} for an example). Instead of a single column with the $\pm$
marks, multiple columns are included, one per merge parent, allowing concise
description of which lines differ against which parents. Furthermore, by
excluding hunks that contain only changes coming from one of the parents, we
get a diff describing only changes actually introduced by the merge itself ---
manual resolutions of merge conflicts.\footnote{Conflicts are further
explained in \ref{conflict}.}

\subsubsection{VCDIFF (RFC3284)}
\label{vcdiff}

The VCDIFF format \cite{vcdiff} deserves a brief mention,
being the official standardized way to exchange binary deltas over the network,
being specified by RFC3284.
However, to our knowledge no widely used version control system actually uses
this format for anything, though it is the default output generated by the
xdelta tool (\ref{xdelta}). The main reason is that it is easier and
faster for individual systems to use their native delta formats even for data
interchange and that the VCDIFF format is rather baroque, implementing it is
not trivial and it is generally not as compact as the native protocols.

\subsubsection{RCS Delta}
\label{rcsdelta}

RCS uses the simplest representation of changes sequence that also spilled over
to CVS and heavily influenced the SVN BerkeleyDB schema.\footnote{SVN FSFS
format is quite different and perhaps most resembles the Mercurial revlogs
(\ref{revlogs}). We chose not to describe SVN formats in detail since we deem
them not particularly interesting in themselves.} \cite{RCS} The representation
is in text format and is optimized for quick access to the latest revision,
which is stored at the top of the file and always in full. Older revisions are
then stored each as a delta against the succeeding one. In case of branches,
the delta direction is reversed and newer revisions are represented by forward
delta --- thus, to get a tip of a non-trunk branch, delta sequence all the way from
the latest trunk revision to branch base and then back to the branch tip
must be applied.

The delta itself is simply composed by lines each containing an add $a$ or
delete $d$ command, followed by line number in base data and number of lines,
with the lines to insert by $a$ following the command immediately.

Aside of very slow reconstruction of non-trunk branch tips and old revisions in
general, the major issue with this format is the requirement to rewrite the whole
file in case of commit, making the repository database much prone to lock
contention and data corruption and making creation of new revision quite an
expensive operation.

\subsubsection{Weave}
\label{weave}

The weave format introduced by SCCS \cite{SCCS} used to be rather obscure for
long time, but currently many version control developers are familiar with it
due to its prominent usage by BitKeeper and some newly developed merging
algorithms.  Still, it has not seen widespread adoption.\footnote{The Bazaar
project actually used it for brief period but then switched to a more traditional delta
chain representation for the benefits of append-only data structure.
\cite{bzrweave}}

Instead of storing each revision separately, this format intersperses all
revisions, listing all the lines that ever appeared in the file together with
the list of revisions they appear in. Then, retrieving any revision means
simply extracting all the lines that have the revision in their set.

This format shares many disadvantages with the RCS delta format, especially
with regard to locking and data corruption potential. In addition, retrieving
any revision is uniformly proportional to history size. On the other hand, the
line annotation\footnote{Information about who, when and in which revision
added a particular line.} is available very cheaply (compared to systems using
delta-based storage where retrieving this involves walking all the history) and
having per-line history information enables some interesting merging methods to
be used (see \ref{pcdv}).

\subsection{Repository Database Formats}

\subsubsection{Trivial Approach}

The trivial approach is to simply use separate files for appropriately
fine-grained objects. This usage is very common in many systems; it is the
approach taken by CVS and Mercurial where every file in the project directory
has its history stored in corresponding file in the repository. It is also the
basic storage format of Git, where each object (commit, tree or blob ---
particular state of a version-tracked file) is stored in a spearate file.

\subsubsection{Git Packs}
\label{gitpacks}

As mentioned above, Git simply stores its first-class objects directly on the
disk as its primary format. However, as explained earlier this is very
inefficient for long-term storage; for that purpose, Git Packs offer
a storage format showing extreme size effectivity in practice.\footnote{Git
also uses the same format for native network data transfer.}

A Git Pack \cite{gitmanual} consists of a pack file and index file. The index
file provides a simple two-level index of the objects keyed by the object id.
The pack file contains the compressed Git objects of all basic types --- {\tt
commit}, {\tt tree} and {\tt blob} --- but can also accomodate an extra object
type {\tt delta}:

The pack-specific delta objects represent a given object by describing binary
differences against another object of the same type,\footnote{Or, of course,
another delta object. However, the delta chain is upper-bounded, by default by
the value of 10.} as yielded by the xdelta algorithm (see \ref{xdelta}).
In theory, any object of the same type will do as a base, but Git sorts the
objects using a special heuristics and then considers $n$ following objects in the list as
delta candidates,\footnote{By default,
the delta window is $n=10$, but commonly, values like 50 and rarely even 100 are
used for generating ultra-dense packfiles.} picking the smallest delta generated.

The list heuristics is the core of the Git Pack
efficiency. \cite{gitpack} \cite{gitsrcdiffdelta} The important criteria are that
blob objects are sorted primarily by the filename they were first reached
though, with the last characters being the most significant ones.  (Thus, all
{\tt .c} files are grouped together, then all files with the same name in
different directories, etc.) The secondary sorting order is by object size,
largest objects first --- thus, deltas are generated from larger objects to
smaller ones, allowing more efficient representation.

Note that the delta order is in almost no way related to physical ordering of
the objects in the pack file.  There are no theoretical restrictions on that,
however Git orders them based on breadth-first search on the graph of all
objects starting from the top commit objects on the defined branches. In other
words, the objects necessary for recreating the latest commits are lumped
together at the beginning of the pack file. This particular ordering greatly
optimizes the I/O operations locality for the most common I/O patterns.

One specific exception in the physical ordering is that in case of delta
objects, the base object is stored before the delta object; this ensures
locality when reconstructing delta chains. In practice, this rule does not
reorder the objects significantly --- as the {\it Linus' Law} states: ``Files
grow'' \cite{gitpack}; thus, the base objects (larger files) tend to come
before delta objects (smaller files) anyway and while it is not enforced, the
delta objects tend to become ``backward deltas'' most of the time.

Git Packs use no locking mechanism --- they are not created on the fly,
but only at certain points in time during a ``garbage collection'' operation
over the repository, from loose one-per-object files created in the repository
by regular usage.\footnote{One exception is when new commits are pulled
by the native protocol, which itself is in the Git Pack format; the data
is then saved directly as a pack.}

\subsubsection{Mercurial Revlogs}
\label{revlogs}

Another popular data structure are so-called revlogs used by Mercurial.
\cite{revlog} \cite{hgsrcrevlog} While Git packs are designed around
heuristics, guarantee very little about worst-case performance and are
optimized for work with hot cache,\footnote{That is, when the user uses the
system continuously for longer time and most of the relevant data is cached in
memory.} revlogs can guarantee good worst-case seek performance (at the cost of
less flexibility for improving average performance) and are tailored for cold
cache access.\footnote{With ``cold cache'', most of the data are still on
disk.} Unfortunately, while there are some casual benchmarks available on
the web, detailed comparison with clearly defined cache status to confirm
these expectations is yet to be published.

Mercurial uses separate revlogs for each tracked file and two extra revlogs for
the manifests and changesets. Each revlog consists of an index file for
revision lookup\footnote{A slightly modified RevlogNG format used by newer
Mercurial versions supports interleaving of index information and data hunks.}
and data file containing a hunk per revision, every hunk being either a full
revision in verbatim or a delta against the previous revision --- the total
size of the delta chain is limited by a small multiple of the revision
size.\footnote{Mercurial 1.0 uses the value of $2$.} In case of the revision
being represented by a delta, the index will contain pointer to the base
revision so that all the necessary data can be read from the file in a single
swipe. Revisions from different branches are all stored in a single revlog, but
when forming delta chains only the order of recording the revisions matters,
not their branch.

When recording new revision in a revlog, the revlog is locked for writing and a
new hunk is appended to the data file and then recorded in the index.  Since
the revlogs are append-only, no locking is required for readers and it is a
matter of simply truncating the file to recover from an interrupted write.
Furthermore, when working with the whole tree (both reading and writing), all
the revlog files are visited in a fixed order compatible with the standard
system files ordering; this way, the filesystem can maintain the most desired
on-disk layout and using standard system tools to copy a repository will result
in a good on-disk layout as well.

\section{User Interaction}

In the development of new generation version control tools, the importance of
good user interface has been underestimated in the past, while it tends to be
one of the most important factors for users when deciding which system to use;
good user interface is therefore one of main considerations in the development
of currently widely used tools.

Traditionally, all of the tools provide command-line interface, which tends to
be the main way of interaction with the system from the side of the developers.
However, many authors also explore ways to interact with the system efficiently
using GUI, both for actually operating the system and merely exploring the
history. Another separate area are web interfaces for inspection of repositories.

\subsection{Command-line Interfaces}

One important conisderation with the command-line interfaces is to keep in mind
that users are actually going to {\sl type} many commands --- e.g., GNU Arch
requires usage of extremely long revision identifiers, and for systems using
hashes for object identification, it is important to allow shortcutting
mentioned earlier.

Another considerable point is to design the interface carefully and logically
around a minimal set of commands, also hiding implementation details whenever
possible (e.g., by trying to intelligently default to a suitable merge algorithm
instead of requiring the user to decide on one). Git provides currently about 140
commands, while large portion of them is in fact almost never interesting to
end users and is useful only as helper commands within scripts or other
commands.\footnote{To simplify the usage of Git, we created a frontend Cogito
that used lowlevel Git interface to create an easy-to-use version control tool
carefully designed to simplify many workflows and present a simple and consistent
set of commands to the user, with consideration for users used to older
systems like CVS or Subversion. Since most features provided by Cogito have
been adopted or obsoleted by upstream Git development, Cogito is not maintained
anymore (but we still consider the Git UI somewhat lacking).}

Note that even regarding traditional command-line interfaces, there can be
interesting variations --- for example, Git extends the basic edit--commit
workflow with the optional {\it Git index} concept, presenting a staging area
inbetween the working tree and next commit; see \ref{gitindex} for details.

User interface can take very unusual forms; a specific one is integration of
version control within the filesystem interface. In that case, accessing a
particular revision of a file is usually performed by simply opening a
specifically crafted interface. This concept goes as far back as to the {\bf
VMS} operating system where the filesystem interface had builtin support for
file versioning. \cite{vms} A configuration management system with version
control capabilities {\bf Vesta} provides primarily filesystem-based access to
individual revisions (furthermore with $O(1)$ access guarantees). \cite{vesta}
There also exist alternative filesystem-based interfaces for more traditional
version control systems as well (though mostly in early development stages),
e.g.  {\bf Bzr-FS} \cite{bzrfs} or {\bf GitFS} \cite{gitfs}.

\subsection{Basic GUI Tools}

There are several types of GUI tools. The most obvious candidates are all-in
tools covering most of the system functionality, often reusing other helper
applications (such as history browsers) --- most systems have tools like that,
e.g. {\bf WinCVS}, {\bf git-gui}, {\bf Bzr-Gtk}, etc.

An interesting approach is to integrate the version control system with basic
system shells or file managers (like the Windows Explorer or Konqueror);
version-controlled files include the state in their icon (clean, modified,
containing merge conflicts, ...) and all versioning operations are available
through the context menu. The classical example being {\bf TortoiseCVS}, clones
like {\bf TortoiseSVN}, {\bf TortoiseBzr}, {\bf TortoiseGit}, {\bf TortoiseHg},
etc. also exist.
Similarly, version control capabilities are often integrated with IDEs
like Eclipse.

Another family of tools are simple GUI helpers designed merely to be used in
concert with the classical commandline controls; this has been mostly
pioneered by BitKeeper, providing citool, mergetool etc. to allow to perform
operations like reviewing to-be-committed state or merging files more
comfortably. Currently, e.g. Git frontends like git-gui or tig provide a way to
control to-be-committed content down to the per-hunk level of diffs and there
exist system-agnostic tools for graphical review of diffs ({\bf xxdiff}, {\bf
kdiff3}, ...) and even merging assistance (especially {\bf meld}).

Visual UI tools do not need to be always graphical. Many developers prefer
exclusively terminal work and sometimes it may be necessary to work on a remote
server over ssh session; {\bf tig} is a good example of terminal tool that
provides ASCII-art\footnote{Graphics constructed in text mode from standard
letters and symbols.} history diagrams and frontends many of Git's functionality
(including index management down to diff-hunk level), without requiring any
graphics rendering.

\subsection{History Browsing}

With the advent of distributed version control, it became much more important
to properly visualize the project history --- suddenly it is not a simple tree of
commits, but a more complex graph with possibly many branchings and merges and
the relationships between revisions may not be immediately obvious to the user.

The first popular history visualizer were BitKeeper's revtool and {\bf
Monotone-viz} \cite{mtviz} which focuses
on visualising the revision graph, offering it as the main visual object to the
user; clicking on nodes will display revision details. During Git infancy, the
tool was adopted as Git-viz, but it did not see widespread usage; however,
similar tool is being used in Bazaar as part of BzrTools.

Instead, a much denser history presentation gained popularity in the Git
community within a tool {\bf gitk}. Its primary visual objects are the
revisions themselves, with the first line of the commit message (traditionally
regarded as kind of ``subject'') and authorship information shown for each
commit; the graph is still shown, however only in narrow area at the margin of
the screen, with nodes aligned to the line structure of commits. Reused in
other Git user interfaces (both graphical ones and tig), this way of
visualization became popular in the Mercurial community as well ({\bf hgk}),
Monotone even produces similar kind of output in ASCII art with its native {\tt
mtn log} command; there also exist projects like {\bf git-browser} which use
AJAX technologies to bring this interface to the web.

\subsection{Web Interfaces}

Important part of user interface of a version control system is a web interface
--- most projects have some kind of web interface available so that people
wanting to look at the source and the history do not have to install the
particular version control tool and download the whole project. Good interface
also should not assume users' familiarity with the particular system, since it
gets much more diverse audience than other tools of the system.

Historically, {\bf cvsweb}, {\bf viewcvs} and similar tools for SVN focused on
primarily presenting the tree of files, with support for inspecting individual
history of files when needed. While similar approaches are also present in some
web interfaces for distributed version control systems (e.g. {\bf GitHub}),
the paradigm shift to
primarily presenting the project history is visible here as well, in the widely
used {\bf gitweb} interface and the {\bf hgweb} clone.  On the project front
page, the user is not presented with root directory of the file tree, but
instead with the summary of the latest commits, available branches and tags,
etc. --- browsing the tree is of course possible, but not the primary activity
anymore.

Something of a rarity, the {\bf Gist} service\footnote{Part of the GitHub cloud.}
uses modern web technologies
to allow users build whole repositories from scratch from within the web
browser; the user is presented by a text-area where they can paste content
(akin to various {\it pastebin} services), but further modifications are then
stored in git revision history, it is possible to add multiple files and
eventually pull the history. It is unclear to us how often it is actually
used as anything more than a trivial one-off pastebin, however.

\section{Merging Methods}
\label{merging}

One of the major past challenges with systems encouraging massive branching was
devising a powerful merging mechanism that would make the merging process
sufficiently easy, quick and robust for frequent usage. This ability is
crucial, otherwise people will become reluctant to create branches and the main
strength of distributed version control will dwindle. On the other hand, practice
shows that merge algorithms should be simple enough to be easily predictable,
even if the tradeoff would be to require manual interventions by the user more
frequently --- common problem of algorithms not based on three-way merge is that
while they may generate less conflicts, their merge resolutions may be relatively
unintuitive and some of their properties are controversial and unfamiliar to the
users.

In this section, we will describe the most common merge strategies
({\it three-way merge} and {\it recursive merge}) as well as some more
advanced theoretical developments ({\it mark-merge}, {\it PseudoCDV merge}).
We will also consider the {\it patch algebra} based way of merging
in changeset-oriented systems.

There is a great number of considerations when designing and comparing merging
algorithms and we cannot cover them all here; we recommend the pages dedicated
to merging algorithms in the RevCtrl wiki \cite{revctrl} for detailed study
of all the aspects.

\begin{figure}[ht]
\centering
\includegraphics[scale=0.6]{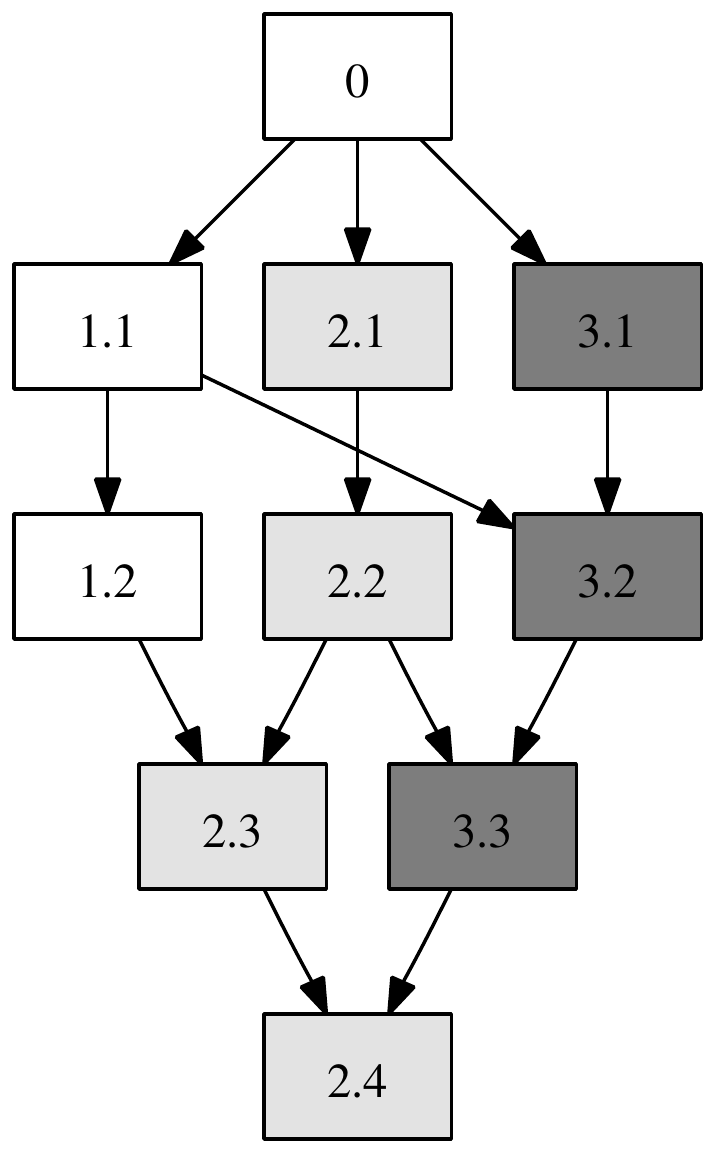}
\caption{A history DAG example}\label{fig:dag}
\end{figure}

\subsection{Fast-forward Merge}
\label{fastforward}

First, we should notice what happens in the revision DAG when a ``blank'' merge
is actually performed: consider merging revisions 1.2 and 2.4 as shown in the
figure \ref{fig:dag}. On the branch 1, no new revisions appeared since its last
merge with 2.2, thus merging the 2.4 revision will effectively merely copy this
revision over to the branch 1. A system might create a merge-node 1.3 anyway,
but this may not be desirable since many useless merge-nodes will be created
in the extremely common scenario of users only tracking some project in read-only
fashion: every time they update their copy, no local revisions will be found
locally but the merge will create a merge-node revision to represent the
update anyway.

To avoid this problem, a ``virtual'' merge strategy of {\it fast-forwarding} is
used; in case one of the merged branches contains no new
revisions since the last merge, it is simply repointed to the head commit of
the other branch.  Thus, in our scenario, branch 1 is simply repointed to the
revision 2.4!  Then, when committing a new revision to branch 1, 2.4 will be
treated as the new fork point.

This technique allows branches of the DAG to converge whenever possible,
however it also breaks commit parents ordering --- after fast-forwarding to a
different branch, merges with the original branch will appear ``flipped''.  The
users either cannot rely on the parents ordering, or they have to constrain the
system usage by a policy to disallow scenarios that would provoke a
fast-forward.\footnote{Fast-forwarding can frequently occur in tight
push-pull loops performed concurrently on a shared central repository.
One way to avoid them is to tell the developers not to merge the changes
their peers have done in parallel and pushed earlier, but instead to
{\it rebase} their own changes on top of the already-pushed ones: this
operation rewrites the history locally, sequencing two commits performed
in parallel.}

\subsection{Three-way Merge}

{\it Three-way merge}\footnote{Unfortunately, we were not able to find out
who introduced the technique of three-way merging, currently commonly known
and widely used; the earliest reference we were able to find is in the
RCS paper. \cite{RCS}} is the most common method of merging in use. Assume
merging two version nodes $x$ and $y$ into merged version $m$. Let $b$ be the
{\it merge base} --- the ``nearest common ancestor'' of $x$ and $y$. Then,
using difference function\footnote{See \ref{delta} for the tour over
various functions.} $\Delta\colon (text1, text2) \to delta$ (and delta applying
function $\Delta^{-1}\colon (text1, delta) \to text2$):

	$$ d_x = \Delta(b, x) $$
	$$ d_y = \Delta(b, y) $$
	$$ d = d_x \cup d_y $$
	$$ m = \Delta^{-1}(b, d) $$

Simply the combination of differences between base and both merged versions are
applied to the base version again and the result is the merged version.  The
problematic part here is determining $d$ --- if $d_x$ and $d_y$ concern
discrete portions of the base version, combining them is trivial.

\label{conflict}
However, when
both change the same part of base version, each in a different way, they cannot
be combined in a simple way. A {\it merge conflict} arises, only the
non-conflicting parts of $d$ are applied and for the rest the user is required
to choose between $m_x = \Delta^{-1}(b, d_x)$ and $m_y = \Delta^{-1}(b, d_y)$
or combine $m_x$ and $m_y$ manually --- usually the system inserts both
variants to the file, separated by visible one-line markers.\footnote{In some
cases, the system can still resolve simple conflicts, for example if they
concern only whitespace characters or perhaps even when merging a simple
reformatting change with a semantic change. Putting more inteligence into
conflict resolution is one of the active areas of merge research. However,
silently mis-merging incompatible changes can have dangerous consequences, so
making the resolution more intelligent at the risk of increasing error rate can
be very harmful.}

In general, in single-shot usage\footnote{When the algorithm is applied on two
arbitrary versions without historical context. In a generic DAG, selecting
inappropriate $b$ within repeated merges can lead to information loss, as
described below.} the algorithm can work on any $b$ merge base version.
However, the choice of a particular $b$ has major impact on the size of
$d_x$ and $d_y$, affecting the conflicting portion of $d$. Thus, $b$ should
be chosen intelligently, usually by taking the {\it least common ancestor} (LCA).

In case of a tree, LCA can be simply described as the node nearest to the tree
root on a non-oriented path from $x$ to $y$, and it will be always unique as
follows from tree properties. However, this approach is satisfying
only for the first-time merge of the branches of $x$ and $y$ --- if a merge
between the branches is performed repeatedly, the same $b$ will always get
chosen, even though the previous merge outcomes would be much more reasonable
choices as the size of $d_x$ and $d_y$ can be assumed to generally increase
over time.

This is the main motivation behind making the history a DAG instead of a tree;
when recording the $m$ version, both $x$ and $y$ are marked as its ancestors
instead of just one of them. When determining the LCA\footnote{In case of a
DAG, we prune paths that are supersets of other paths from our consideration.}
of descendands of $x$ and $y$ later, $m$ can then be used instead of the
original $b$. Thus, for example for the graph in figure \ref{fig:dag} earlier,
the base of the 2.4 merge of 2.3 and 3.3 would be 2.2.

However, in this case we lose the guarantee of $b$ uniqueness, since multiple
discrete paths between $x$ and $y$ can exist: to deal with this problem,
recursive merge extension of the algorithm has been devised.

\begin{figure}[ht]
\centering
\includegraphics[scale=0.6]{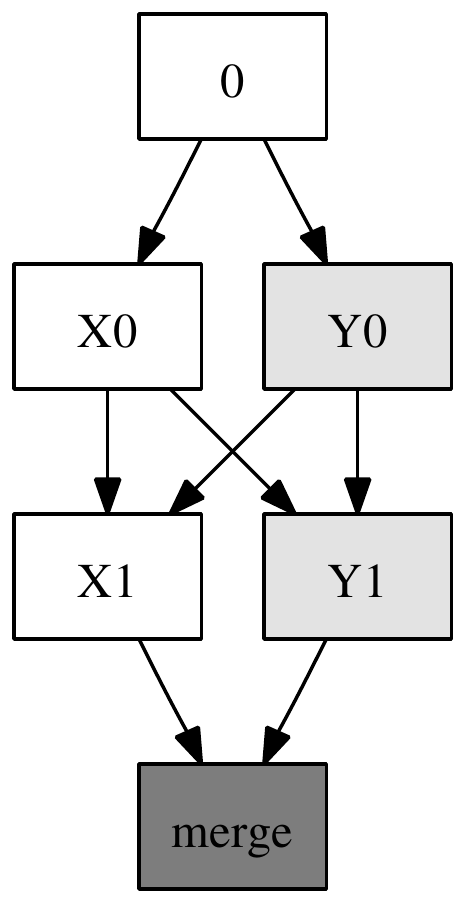}
\caption{A criss-cross history graph example}\label{fig:crisscross}
\end{figure}

\subsection{Recursive Merge}

Consider the {\it criss-cross merge} scenario \cite{crisscross} shown in
figure \ref{fig:crisscross}.
In this case, $b$ can take
the values of both $X_0$ and $Y_0$, however taking any of these can lose data
from the other branch depending on the nature of the merge --- consider
conflicting change in $X_0$ and $Y_0$, in $X_1$ resolved by taking $X_0$
version exclusively and in $Y_1$ by taking $Y_0$. Then, for $b = X_0$, $Y_0$
version will be cleanly merged and vice versa --- clearly, conflict should
be generated instead.

This particular problem and its more complicated variations led some
researchers to abandon the three-way merge altogether and focus on other merge
methods instead, some of them described below. However, the problem can be at
least partially solved within the context of three-way merge as well, though
the practical impact has to be yet evaluated carefully.

A basic rule should be that the system, detecting multiple possible merge
bases, does never randomly select one but instead asks the user for intervention,
chooses a proper merge base or possibly intelligently constructs one.

One naive approach we devised is to take ${\rm LCA}(X_1, Y_1)$ recursively and repeat
until a single revision comes out. The practicality of this approach is
dependent on the development strategy adopted by a particular project --- if
the criss-cross merge is not repeated for too long and a common base for both
branches appears once in a while, the LCA will stay in a reasonable distance
from the merged versions and conflicts would generally stay within reasonable
scale. However, for long-term criss-cross merges this approach degenerates to
something similar to a tree-based history, and it still
leaves some more advanced criss-cross problems unsolved
\cite{arcrisscross}.\footnote{It must be also considered that if the merge
generates unnecessarily large conflicts, risk of human-induced mismerge raises
considerably.}

An important enhancement implemented by Fredrik N. Kuivinen is to actually
perform the merges themselves recursively: \cite{recmerge} \cite{crisscross}

	$$ |B| = 1: \quad b(B) = B_0 $$
	$$ |B| = 2: \quad b(B) = M({\rm LCA}(B_0, B_1), B_0, B_1) $$
	$$ M(B, x, y) = \Delta^{-1}(b(B), x \cup y) $$
	$$ m(x, y) = M({\rm LCA}(x, y), x, y) $$

That is, in the example above:

	$$ m = \Delta^{-1}(\Delta^{-1}(0, X_0 \cup Y_0), X_1 \cup Y_1) $$

It is easily visible that this works properly in case of no conflicts, and
reduces conflict rate by working on much finer grain level. In case of
conflict, the main idea of the algorithm is to simply {\sl leave the conflict
markers in place} when using the result as a base for further merges. This
means that earlier conflicts are properly propagated as well as conflicting
changes in newer revisions. However, some rare edge cases are still mishandled
as described in \cite{crisscross}.

\subsection{Precise Codeville Merge}
\label{pcdv}

A weave-based merge algorithm {\it Precise Codeville Merge} \cite{pcdv} has
been devised by the {\bf Codeville} version control system.\footnote{The system is
not otherwise covered here since it has never seen wider usage and is not
developed anymore; it can be probably said that its main purpose ended up to be
to research various merging algorithms.} It later turned out that Codeville
probably independently invented a merging method very similar to what BitKeeper
uses internally.

Instead of comparing the to-be-merged
revisions to a base revision, the algorithm directly performs a two-way merge
between the two revisions and then uses the weave-embedded history information
to decide which side of each conflicting hunk is to be chosen.

Consider $v_i(r)$ being the number of state changes (addition or removal; also
called {\it generation count}) of
line $i$ at revision $r$.  For each revision, have a list of line chunks: $l_x$,
$l_y$.  Then, the weave of the file is iterated line by
line:

\begin{itemize}
\item if a line is not in either of the revisions, nothing is done
\item if a line is found to be only in one of the revisions, it is
appended to appropriate line chunk list $l_x$ or $l_y$ and
revision precedence flag $p_x$ is set if $v_i(x) > v_i(y)$, or
$p_y$ if $v_i(y) > v_i(x)$ (obviously, $v_i(x) \ne v_i(y)$)
\item if a line is found to be in both revisions, either:
\begin{enumerate}
\item $p_x$ and $p_y$ are unset and this is part of common lines block and appended
to the output
\item only $p_x$ or $p_y$ is set and this is non-conflicting
change (then, $l_x$ or $l_y$ respectively is appended)
\item both are set and a conflict block from $l_x$ and $l_y$ chunks is generated;
in the latter two cases, $l_x \leftarrow l_y \leftarrow \emptyset$ and
$p_x \leftarrow p_y \leftarrow 0$.
\end{enumerate}
\end{itemize}

The advantage of this algorithm is the absence of the need to walk the revision
graph at all (provided that the system uses the right storage format, which can
be difficult while keeping the desirable append-only properties of the most
common ones), simplicity and handling of most of the problematic edge cases of
the three-way merges. Also, this algorithm handles cherry-picking naturally
well compared to three-way merge.

The disadvantage is a requirement for the weave structure potentially expensive
to emulate if the system uses different storage method,\footnote{Vesta
currently uses this merge method and constructs weaves on-demand; scalability
of this method was not studied, however. \cite{xorianc}} and somewhat
controversial usage of the $v_i$ criterion --- it is not immediately obvious
that lines with more state changes should always win against these with less.
Due to the fundamental difference in operation to the much more widely used
three-way merge, in cases it chooses different result its operation may be
confusing to the users.

\subsection{Mark Merge}
\label{markmerge}

The mark merge (or *-merge) algorithm \cite{markmerge} is somewhat unusual.
First, all the other merge techniques described here are designed to merge
vectors (file content), while this one is a {\it scalar merge} algorithm
(for example
for directory items --- the executability flag or file name when merging
renames).  Second, mark merge has a precisely specified user model and formal
proof that the algorithm matches the user model.

The aim of mark merge is to define the most sensible way of merging scalar
values while giving foremost priority to explicit user choices.  First, a
{\it mark} means explicit decision by the user on the scalar value; marked
revision is such that user explicitly set the value in this revision.  Let us
have version $r$, then $*(r)$ shall be set of minimal marked ancestors of $r$
--- that is, taking graph of $r$ ancestry and reducing it to marked revisions,
set of revisions with no descendants.  Then, when merging versions $x$ and $y$
with values $v_x$ and $v_y$:

\begin{itemize}
\item if $v_x = v_y$, the value shall be also the merge result
\item if all $(x)*$ revisions are ancestors of $y$, $v_y$ shall be merge result
\item if all $(y)*$ revisions are ancestors of $x$, $v_x$ shall be merge result
\item otherwise, throw a conflict
\end{itemize}

For detailed formal user model and proof of correctness please refer to
\cite{markmerge}.

\subsection{Patch Algebra}
\label{palg}

In Darcs (\ref{darcs}), a particularly elegant formal system {\it patch algebra}
has been developed for merging two sets of patches in a changeset-oriented
version control system. \cite{darcsman}

The main idea of patch algebra is to provide natural and formally-backed
semantics for merging within such a system.\footnote{But please note that while
Darcs makes an attempt for formal description, it is still lacking in many
aspects both regarding precise definitions and many proofs missing.  We must
however mention a relatively little-known theoretical work \cite{principled}
inspired by Darcs formalism and attempting to rebuild it on sound foundations.}
All patches are defined to have a set of dependencies on other
patches\footnote{The dependencies are either logical as explicitly specified by
the user, or syntactic --- a patch is depending on all patches lines of which
it touches --- as autodetected by the system.}, inversions, and independent
patches are allowed to {\it commute} on the stack. When merging two sets, independent
patches are appropriately commuted (sometimes requiring use of inversion
patches to work around fuzzy applications) and for conflicting patches, user
resolution is requested; the merge operation is then recorded in a special
merger patch.\footnote{The actual mechanisms used to preserve all the desirable
properties of the patch stack are rather complicated --- please refer to
\cite{darcsman} for full description.}

Given the way the to-be-merged merger patches are unwound, it is believed that
the merge behaviour with regard to the pathological cases like criss-cross
merge are similar to the recursive merge algorithm or better, since Darcs can
make use even of intermediate changes between the two merged revisions and
their base that the three-way merge approach cannot see. \cite{crisscross}
\cite{badmerge} However, we are not aware of any work formally analysing and
rigorously comparing the properties of Darcs merges with the other approaches.

\section{Future Research}
\label{future}

On the technical side, some of the existing systems are not portable enough to
be usable equally well in different operating systems (especially with regard
to Microsoft Windows vs. Linux and other UNIX systems) and the user interfaces
are still lacking in many aspects, as well as documentation.

Regarding the theoretical side, we believe three main areas need research most
urgently --- content movement tracking, management of third party changes
and more precise formal specification.

\subsection{Fine-grained Content Movement Tracking}
\label{futuremove}

More fine-grained tracking of changes is required to improve merging
algorithms, accounting for changes and following history of arbitrary parts of
content; as noted in \ref{moves}, some of the systems provide way to follow
file history across renames, {\bf Git} introduces the {\it pickaxe} mechanism,
but better infrastructure is required for practical usage especially during
merging and allowing visual representation and usage.

\subsection{Third-party Changes Tracking}

The current distributed architectures still are not flexible enough to
accomodate for fully distributed development in open-source projects. When a
third-party developer publishes some changes and the upstream developers want
to merge it only with certain reservations, the third-party developer can either do the
required changes on top of his previous changes, leaving the history in a
certain degree of disarray and making his proposed chances increasingly
harder to review, or rewrite the history of his changes, which creates
problems for others who have already accessed the published version
since the rewrite essentially created a fresh new branch unrelated to the
older one.

Similar problems arise when the third-party changes need to be ported to new
upstream version, or when a distributor wants to maintain series of changes in
their version of the product; using the traditional merge mechanisms, it
quickly becomes difficult to track the changes against upstream cleanly as the
ability to generate a diff of single change against latest upstream version can
be crucial.

Thus, many people make use of the distributed nature by only maintaining the
changes locally and still using patches for changes interchange; or, they use
the version control system for tracking upstream and some special frontend on
top of that\footnote{E.g. {\bf mq} for Mercurial, {\bf StGIT} for Git} for
tracking their changes in form of stack --- this has user interface issues and
it is problematic to publish the repositories since the history is changing
constantly. Another alternative (often used by open-source software
distributors when maintaining their packages) is to version-control the patches
in their diff form directly.

Using patches for changes interchange can have some benefits with regard e.g.
to code review practices, but frequently it is merely extra overhead and it
incurs extra load on the third-party developers.  A new design that would allow
seamless external changes tracking needs to be developed.

\subsubsection{TopGit}

We have created {\bf TopGit} \cite{topgit}, a layer over Git
that aims to solve the task of maintaining third-party patches properly,
allowing for fully distributed development and proper history tracking
of all the changes.

We start off from the concept of {\it topic branches}, popular Git technique
of having a specialized branch for development of each independent change,
then merging them all together; this is merely a ``design pattern'', needing
no special tool support. However, we extend this in two ways:

\begin{enumerate}
\item We create a directed acyclic graph from the topic branches,\footnote{Thus,
conceptually one level higher than the DAG of the {\sl commits}.}
giving each branch a list of dependencies (other branches).
\item We store metadata for each topic branch within its file tree:
the branch description and authorship information within {\tt /.topmsg}
file and the list of branches it depends on inside {\tt /.topdeps}.\footnote{These
two files are excluded from merges of dependencies.}
\end{enumerate}

The topic branches can depend on other topic branches, but also on regular Git
branches. Typically, in the main set of third-party patches, each patch will
have its own TopGit branch depending on a Git branch tracking the upstream
development, then possibly some third-party patches will have dependencies
on some of the TopGit branches if they further extend the third-party changes,
and at the top of the graph will be usually at least one so-called
{\it staging branch} that introduces no changes on its own (as opposed to
{\it content branches}) but ties all the third-party patches together
by listing them all in its dependencies.

TopGit provides tools for creating topic branches, querying them, pulling
topic branch updates and merging them with local work, but most importantly
a command for updating the branches, which recursively updates all the
dependencies of a branch, then merges them in.

The project reached basic practical usability and at the time of this writing
is used for maintenance of some Debian packages, but its
user interface leaves a lot to be desired and some of the functionality is not
fully implemented.

Especially problematic is the task of {\sl removing}
a dependency of a patch in a way that preserves the history, makes it possible
to easily re-add the dependency later and does not spread the dependency
removal to further depending patches that also have the removed patch as
a dependency. \cite{topgit-removal}
Another problem is inventing a user interface that makes history exploration
easy, since traditional commit browsers present exceedingly complex trees.

\subsection{Formal Analysis}

Precise semantics of current version control systems are defined only loosely
and the lack of formal analysis is visible especially in the area of merging
algorithms. In order to get a truly reliable and dependable system with regards
to merging ``wild'' branches exhibiting pathological behaviour, the precise
semantics of various merges should be specified formally; many of the currently
widely used merging methods lack detailed formal analysis and their behaviour
in corner cases is only guessed.

There has been ongoing development in this area \cite{darcsman}
\cite{principled}, but it appears to be mostly stalled now as the recursive
three-way merge and similar algorithms appear to work well enough in common
cases and thus there is not much incentive to formally describe their behaviour
in corner cases. However, we believe that such a formal foundation would
also help development of newer changeset-oriented systems, in turn
making progress with the third-party changes tracking problem.

\section{Conclusion}

We have gone through the most important concepts, algorithms and structures
used in current version control systems, and put them in the context
of practical usage; we hope this gave the reader a comprehensive idea
about the area and will help to put his further research in concrete
topics into the broad context.

Among the open source projects, the new generation of distributed version
control systems has registered massive uptake; while there are still projects
using CVS and many projects are using or even migrating to SVN, large portion
of the most high-profile open source projects with many commits per day and
code base of millions of lines of code \cite{sloc} is using one of the
distributed systems described above.

The wide adoption and scale of use clearly indicates that the current systems
have reached appropriate scalability and reliability levels. The high number of user
interface tools in development also makes it more comfortable to use these
systems and smoothens up the learning curve that has been traditionally very
steep for distributed version control systems. However, as described in section
\ref{future}, many desirable features are still unimplemented and support for
some workflows is still very lacking.

\subsection{Acknowledgements}

This article is based on Bachelor thesis I wrote at the Faculty of Math and
Physics, Charles University. I would like to thank my thesis advisor
Martin Mareš for
guidance and Bram Cohen, Piet Delport, Junio C. Hamano, Greg Hudson, Uwe
Kleine-König, Tom Lord,
Matt Mackall, Larry McVoy, Kenneth C. Schalk, Nathaniel Smith, Pavel Surynek
and Zooko Wilcox--O'Hearn for their review and helpful input.


\end{document}